# Noise activated dissociation of soft elastic contacts

M. K. Chaudhury[*] and P. S. Goohpattader

Department of Chemical Engineering, Lehigh University, Bethlehem, PA 18015, USA

## Abstract

Adhesive forces are capable of deforming a soft elastic object when it comes in contact with a flat rigid substrate. The contact is in stable equilibrium if the total energy of the system arising from the elastic and surface forces exhibits a minimum at a zero or at a slightly negative load. However, as the system is continually unloaded, the energy barrier decreases and it eventually disappears, thus leading to a ballistic separation of the contact. While this type of contact splitting has received wide recognition, what has not been much appreciated with these types of soft adhesion problems is that rupture of a contact can also occur at any finite sub critical load in the presence of a noise. The soft contact problems are unique in that the noise can be athermal, whereas the metastable and stable states of the thermodynamic potential can arise from the competition of the elastic and the interfacial energies of the system. Analysis based on Kramers' theory and simulations based on Langevin dynamics show that the contact rupture dynamics is amenable to an Eyring's form of a force and noise induced escape of a particle from a potential well that is generic to various types of colloidal and macromolecular processes. These ideas are useful in understanding the results of a recent experiment involving the noise activated rolling dynamics of a rigid sphere on a surface, where it is pinned by soft micro-fibrillar contacts.

[*] e-mail: mkc4@lehigh.edu

# 1 Introduction

Morphological and/or elastic heterogeneities can play important roles in improving the toughness of an adhesive interface [1-3]. Built upon the path breaking ideas of Thomson et al [4, 5] and Kendall [6], it is now well appreciated that such heterogeneities are capable of trapping a crack locally and intermittently. Every time a crack is de-pinned from such a trapped state, some energy is dissipated; thus the overall fracture toughness is enhanced. Examples of defect enhanced fracture toughness are plenty in natural and laboratory settings, which have been reviewed [1, 3] recently. The main emphasis of the conventional treatments has, however, been on the ballistic separation of surfaces from a pinned state. What has not been much appreciated is that these joints, like all systems in nature, are subjected to various types of noises originating from thermal, environmental, and mechanical processes. It is therefore imperative to develop an understanding of how two surfaces separate from a pinned state in the presence of a noise. The subject of this paper is to illustrate this situation with a specific example of the rolling of a rigid sphere on a surface, where it is initially pinned by deformable elastic fibrils but is de-pinned when it is subjected to a low strength mechanical noise. We discuss the kinetics of such a phenomenon after providing the required backgrounds on how a pinning potential develops from the completion of elastic and surface forces in a soft elastic system.

Beginning with the pioneering works of Johnson, Kendall and Roberts and others [7-18] it is now well-established that the interfacial forces can deform a soft elastic object when it comes into contact with another rigid material. Several studies [15-18] have also pointed out that the adhesive forces can be so significant that a soft object jumps into contact with another material when they are in close proximity following which one or both of them may deform elastically.

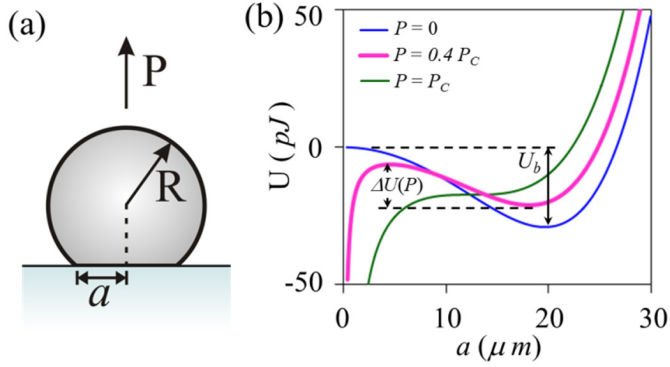

**Fig. 1**. (Colour on-line) (a) Schematic of a sphere in contact with a flat substrate. A negative load (P) is applied on the sphere of radius $R$ and contact modulus of $E^*$. (b) Total energy of the system at fixed loads but at different values of the contact radius calculated with the following parameters. $R= 100$ $\mu$m, $E^*= 1$MPa, $W=0.04$ J/m$^2$. For this combination of material parameters, the critical load $P_c$ is -19 $\mu$N. In the absence of the load, the system has one minimum. However, as the load is increased, a maximum and a minimum appear in the energy potential. At a critical negative load, the energy barrier disappears.

The simplest illustrative case [7, 8, 10] is the deformation of a sphere is shown schematically in fig. (1). For the purpose of illustration, we consider that a negative load ($P < 0$) is applied onto the sphere. The total energy of the system [7, 8, 10] is the summation of the potential, elastic and adhesion energies:

$$U(a,P) = \frac{E^* a^5}{15R^2} - \frac{P^2}{3E^* a} - \frac{Pa^2}{3R} - \pi W a^2 \qquad (1)$$

Here, $E^*$ is the contact modulus, $P$ is the applied load, $R$ is the radius of the sphere, $W$ is the work of adhesion and $a$ is the radius of the contact area. When $P = 0$, the energy $U(a, P)$ exhibits a minimum with a depth of $U_b = \frac{(6\pi W)^{5/3} R^{4/3}}{10 E^{*2/3}}$ that can be easily deduced by setting the first derivative of the total energy of the system (eq. (1)) to zero. The system is unconditionally stable at this stage. However, with a negative load, the energy landscape changes substantially as shown in fig. (1). A local minimum still exists as long as the load is smaller than a critical value, but now an unstable equilibrium state appears in the energy landscape. There is a difference of energy between the unstable and the stable equilibrium states that disappears only at a critical load thus leading to a ballistic separation of the sphere from the substrate. What we emphasize in this paper is that the sphere can explore various states of the energy landscape (fig. (1)) diffusively in the presence of a noise. When the unstable equilibrium state is crossed, the contact falls apart. Like any chemical kinetics, the frequency of this rupture should follow a Van't Hoff-Arrhenius-Eyring [19] type rate law, which is generic to the force induced dissociation of molecular bonds as witnessed in various types of thermally activated processes such as plastic flow [19, 22], wetting dynamics [23], friction [20-22, 24-30], sub-critical fracture [31] and the dissociation kinetics of single molecules [32, 33], to name a few.

We approach the current problem within the framework of a Smoluchowski-Kramers equation [34, 35], in which two physical parameters are important. The first is the barrier height and second is the frequency at which attempts are made to cross the barrier. Several studies [19-33] have pointed out that an external force reduces the height of any pre-existing energy barrier. To the best of our knowledge, Garg [36] was the first to point out that it is not only the barrier height, but also the pre-exponential frequency factor that changes with the applied load. Afterwards several studies [37-41] used the force modulated frequency and the barrier energy terms in the Kramers equation to simulate the dissociation kinetics of polymer chains with a linear loading rate in the style of Evans and Ritchie [33], as well as Schallamach [24]. The findings of the later studies [38-41] agree with Garg [36] in that the applied force ($f$) modifies the energy barrier as ~ $(1-f/f_c)^{1.5}$, where $f_c$ is the critical force of detachment. Recently, such a scaling has been verified in molecular dynamics simulations as well [42, 43]. The finding of Lacks et al [42] is particularly interesting in that they showed that it is not only the energy barrier, but also the free energy barrier that follows the scaling of $\sim(1-f/f_c)^{1.5}$. In the light of these previous studies, we write the overall frequency of rupture of a soft sphere from a solid substrate (assuming a linear friction) as follows:

$$v = \left[\frac{\tau_L}{2\pi}\omega_1(P)\omega_2(P)\right]\exp\left[-\frac{2\Delta U(P)}{mK\tau_L}\right] \qquad (2)$$

Equation (2) is the celebrated Kramers' equation in the strong friction limit, where $\tau_L$ is the Langevin relaxation time, $\omega_1(P)$ and $\omega_2(P)$ are the frequencies corresponding to the curvature of the energy potential near its maximum and minimum values, $\Delta U(P)$ is the barrier height, $m$ is the mass of the sphere. $K$ (m$^2$/s$^3$) is the strength of a Gaussian white noise, which is defined as $\Gamma^2 \tau_c$, $\Gamma$ (m/s$^2$) being the root mean square acceleration of the noise, and $\tau_c$ (s) is its pulse width. The term $mK\tau_L/2$ of eq. (2) is the surrogate for the kinetic energy ($k_BT$) of a thermal system. The random noise can be thermal in micron scale systems or it can be environmental in macroscopic systems. In a controlled experiment at the laboratory setting, the noise can also be generated with a waveform generator and fed to an oscillator. An accelerometer can be used to estimate the acceleration pulses from which $\Gamma$ can be estimated. Details of these procedures can be found in our previous publications [44, 45].

## 2 Noise induced detachment of the JKR like contact

### 2.1 Spherical contact

The energy barrier and the spring constant needed to estimate the frequency of transition can be obtained from a Taylor series expansion of eq. (1) about a critical point $a_i$ as follows:

$$U - U_0 = a\left(\frac{E^*a^3}{3R^2} + \frac{P^2}{3E^*a^3} - \frac{2P}{3R} - 2\pi W\right)\bigg|_P (a-a_i) + \frac{1}{2}\left(\frac{E^*a^3}{R^2} - \frac{P^2}{E^*a^3}\right)\bigg|_P (a-a_i)^2 + \ldots \qquad (3)$$

Setting the first term of the right hand side of equation (3) to zero, one obtains the classical JKR [3] equation ( eq. (4)) that gives two critical values of the contact radius ($a_i$) - one at the unstable ($a_1$) and the other at the stable ($a_2$) position of the energy landscape.

$$W = \frac{\left(\frac{E^* a_i^3}{R} - P\right)^2}{6\pi E^* a_i^3}$$

(4a)

Or, equivalently:

$$a_i^3 = \frac{R}{E^*}\left[P + 3\pi WR \pm \sqrt{6\pi WRP + (3\pi WR)^2}\right] \quad (4b)$$

The curvatures of the potential (second term of eq. (3)) around these two (stable and unstable) equilibrium points yield the spring constants that can be expressed as:

$$m\omega_i^2(P) = \left|\left(\frac{E^* a_i^3}{R^2} - \frac{P^2}{E^* a_i^3}\right)\right| \quad (5)$$

Now collecting all the terms, the frequency of separation of the sphere from the surface in the presence of a negative load $P$ and a noise of strength $K$ can be expressed in terms of the following form of the Kramers equation:

$$\nu(P) = \left[\frac{\tau_L}{2\pi m}\sqrt{\left(\frac{P^2}{E^* a_1^3} - \frac{E^* a_1^3}{R^2}\right)\left(\frac{E^* a_2^3}{R^2} - \frac{P^2}{E^* a_2^3}\right)}\right]\exp\left[-\frac{2}{mK\tau_L}\left(\frac{E^*(a_2^5 - a_1^5)}{10R^2} + \frac{P^2}{2E^*}\left(\frac{1}{a_2} - \frac{1}{a_1}\right)\right)\right] \quad (6)$$

Note that the term work of adhesion (W) is implicit in equation 6, which has been eliminated by combining equations 1 and 4a in order to obtain a compact form of the exponent.

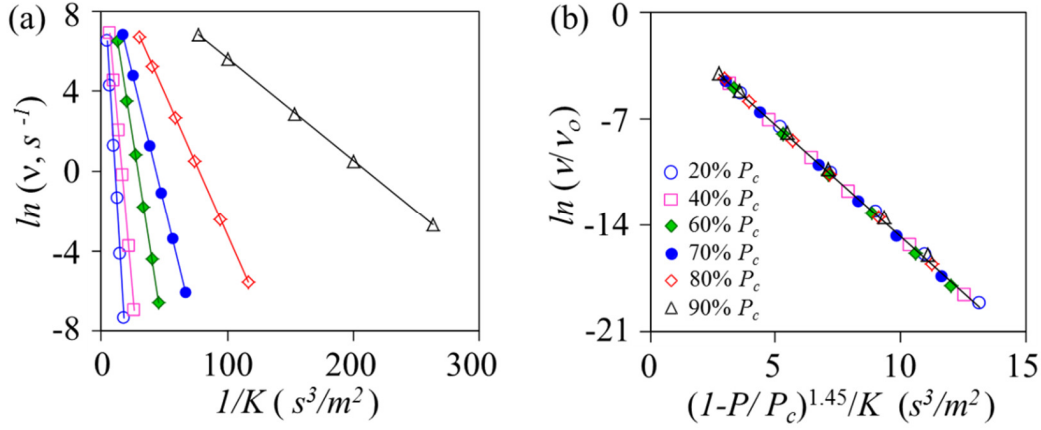

**Fig. 2**. (Colour on-line) (a) Logarithm of the frequency of rupture of a sphere from a flat surface varies linearly with $1/K$ at a given load. These calculations were performed with the following parameters: $R$= 100 $\mu$m, $E^*$= 1MPa, $W$=0.04 J/m$^2$, $m$=4.2 $\mu$g and $\tau_L$=0.01 s. (b) collapse of the rupture kinetic data results when $\ln(v/v_o)$ is plotted against $(1-P/P_c)^{1.45}/K$. Similar symbols in figures (a) and (b) correspond to the same load.

With the reasonable values of the material and geometric properties of a soft elastic contact, eq.(6) was solved numerically. The results, as summarized in fig. (2), show that the logarithm of the rupture frequency is linear with the reciprocal of the noise strength at a fixed value of the applied load that is typical of a Van't Hoff-Arrhenius-Eyring type kinetics. The data obtained at various values of the applied load can also be summarized (fig. (2b)) using eq. (7):

$$v \approx v_0 \exp\left[-\frac{2U_b(1-P/P_c)^{1.45}}{mK\tau_L}\right] \quad (7)$$

Where $U_b = \dfrac{(6\pi W)^{5/3} R^{4/3}}{10 E^{*2/3}}$ is the depth of the potential well in the absence of the load, which we identified earlier in the text. The exponent (1.45) of the reduced bias $(1-P/P_c)$ is close to that (1.5) of Garg's expression [36] and can be verified (approximately) as well by integrating the following form of the rupture dynamics (see Appendix) with a noise term $\gamma(t)$ as follows:

$$\left(\frac{2a}{3R}-\frac{2P}{3E^*a^2}\right)^2 \ddot{a} + \left(\frac{2a}{3R}-\frac{2P}{3E^*a^2}\right)\left(\frac{2}{3R}+\frac{4P}{3E^*a^3}\right)\dot{a}^2 + \frac{E^*a^4}{3mR^2} + \frac{P^2}{3mE^*a^2} - \frac{2Pa}{3mR} - \frac{2\pi Wa}{m} + \frac{4\pi aT\sigma_o^2}{mE^*}\dot{a} = \gamma(t) \quad (8)$$

While the above analysis has been carried out with a circular contact of a spherical object, similar analysis can also be performed with other types of contacts as well. For example, with a flat circular contact [8, 46] with a deformable substrate, the total energy is of the following form:

$$U \approx -\frac{P^2}{3E^*a} - \pi W a^2 \qquad (9)$$

Where $a$ is the radius of contact. For this particular geometry, the barrier energy is:

$$\Delta U(P) \approx W a^2 \left[ 0.89 \left(\frac{P}{P_c}\right)^2 \left(1 - \left(\frac{P_c}{P}\right)^{2/3}\right) + 0.44 \left(1 - \left(\frac{P}{P_c}\right)^{4/3}\right) \right] \qquad (10)$$

Numerical evaluation of eq. (10) leads to a barrier height as $\Delta U(P) \sim W a^2 (1 - P/P_c)^2$. On the other hand, the energy of the contact of a cone [47, 48] of semi angle $\pi/2 - \beta$ is:

$$U(a,P) = \frac{\pi^2 (\sec\beta \tan^2\beta) E^* a^3}{48} - \frac{P^2 \sec\beta}{4E^*a} - \frac{\pi(\sec\beta \tan\beta) P a}{4} - \pi W a^2 \sec\beta \qquad (11)$$

Here, the depth of the potential scales as $W^3/E^{*2}$. Numerical analysis of eq. (11) shows that the force dependent barrier height is of the form: $\Delta U(P) \sim (W^3/E^{*2})(1 - P/P_c)^{1.4}$.

From the above discussions, it is clear that the energy barriers are strong functions of the geometry of the contacting object. While for the sphere and the flat, the energy barrier scales as $W^{5/3} R^{4/3}/E^{*2/3}$ and $W a^2$ respectively, it scales as $W^3/E^{*2}$ for the conical contact that lacks a clear geometric length scale. By contrast, the exponent of the reduced bias is close to 2 for the flat contact, whereas it is close to 1.5 for both the spherical and conical contacts. We now explore how the insights gained from these discussions could be useful to understand certain features of the noise induced micro-fibrillar detachments as we witnessed in our previous studies [44, 45].

## 3 Rolling of a rigid sphere on a fibrillated rubber

Recently, we studied the behavior of the rolling of a small rigid sphere on a low modulus flat rubber that was decorated with the microfibrils of the same material using a lithographic method [49, 50]. A rigid sphere is pinned on such a surface via adhesion to the fibrils. When the substrate is inclined above a critical angle ($\theta_c \sim 2.5^o$), the sphere rolls by de-pinning from the fibrils in the receding edge, but making fresh contact with them at the advancing edge.

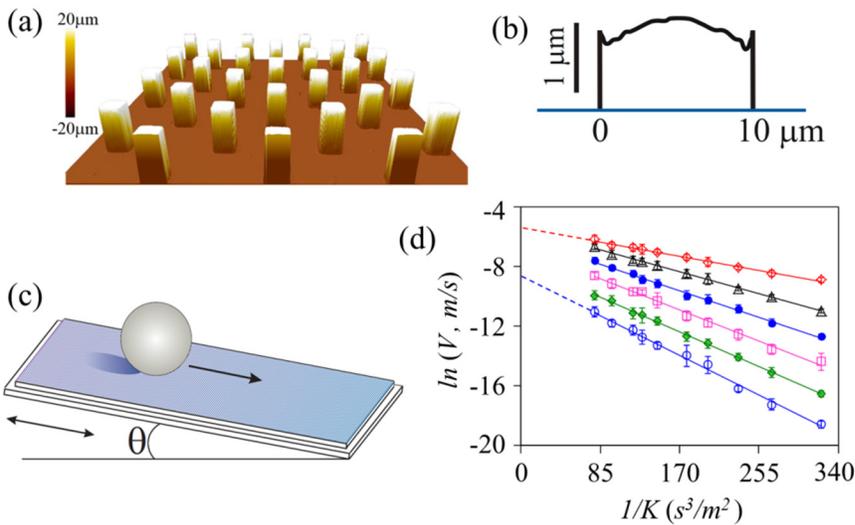

**Fig. 3**. (Colour on-line) (a) 3D Profile of the fibrillar rubber substrate measured with a noncontact optical 3D profilometer (ZeGage with ZeMaps V.1.11, from Zemetrics, Inc.). (b) The profile of the end of a fibril showing that it is slightly curved. The spikes are artifacts arising from the fact that the profilometer failed to follow the edges of the fibrils ( c) Schematic of a rigid sphere (a small steel ball of 4 mm diameter and 0.26 gm mass) on an inclined substrate of a silicone rubber (0.6 mm thick with a modulus of 2.2 MPa), from which square fibrils of the same material are projected outwards on a diagonal square lattice at a spacing of 50 $\mu$m. In the absence of any noise, the sphere rolls at an angle of about 2.5$^o$. However, with an angle less than 2.5$^o$, the sphere rolls with a velocity that increases with both the noise strength and the bias. (d) At each bias, ln($V$) varies linearly with $1/K$ . The symbols are as follows. red open diamond (◊, 0.078mN), black open triangle (Δ, 0.067mN), filled blue circle (●, 0.056mN), open pink square (□, 0.044mN), filled green diamond (♦, 0.033mN), open blue circle (○, 0.022 mN). Some of these data were originally reported in reference [45]. However, in this study, we extended the dynamic range of the noise strength by going to even smaller values of *K*.

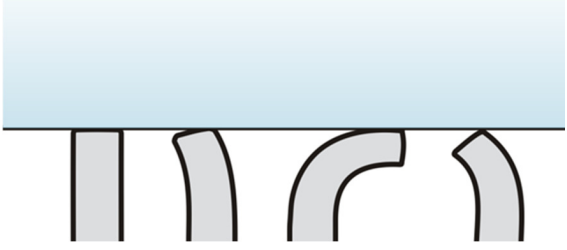

**Fig. 4.** (Colour on-line) Schematic illustrations of the pining and de-pinning events of the fibrils in contact with a rigid sphere.

The ball can also roll sub-critically, i.e. at an angle of inclination $\theta < \theta_c$, provided that it is subjected to an external vibration. In previous publications [44, 45], we reported this type of stochastic rolling behavior of a steel ball on a fibrillated rubber substrate, when the later was vibrated parallel to its base with a Gaussian noise. As discussed in references [45] and [51], the torque applied on the ball by the external force about its point of contact with the surface is balanced by the torque due to adhesion. The contact mechanical force due to adhesion is compressive at the advancing edge of contact, but is tensile at its receding edge. From a balance of the two torques, it can be shown that the collective tension caused by all the fibrils, each experiencing a force of magnitude $P$, is proportional to the applied bias $F$ ($= mgsin\theta$).

The basic observation [45] was that the ball exhibits a stick-roll motion at very low noise strength with the net drift always occurring along the direction of the bias. The rolling velocity of the sphere on the fibrillated rubber could indeed be described by an Arrhenius equation in the sense that $\ln(V)$ is fairly linear with $1/K$ over a substantial dynamic range of the velocity (fig. 3d).

In order to analyze this type of rolling dynamics data in the light of the discussion of section 2, we first need to multiply the fibrillar detachment frequency with a length scale in order to obtain the scale of a velocity. This is, however, not a simple proposition as this length scale itself would depend on how effectively the detached sphere is damped. If the damping is weak, the sphere would roll over several fibrillar spacings before being arrested by another set of fibrils.

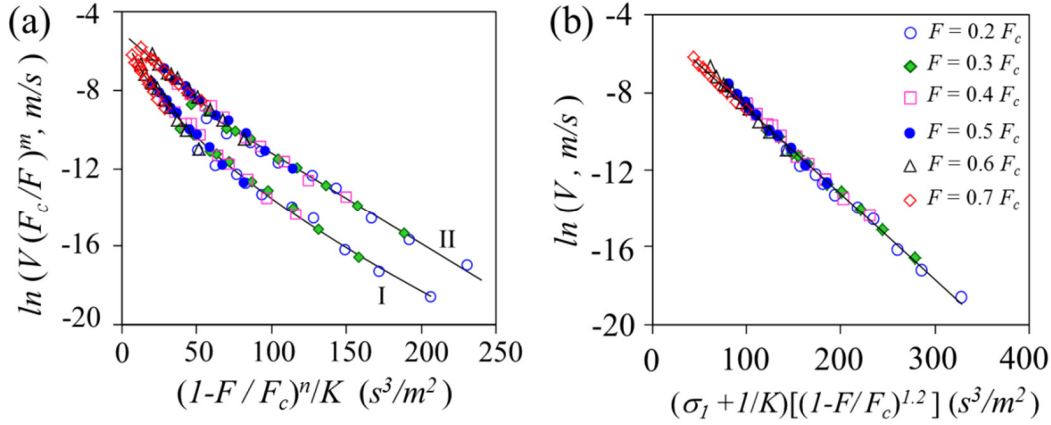

**Fig. 5**. (Colour on-line)(a) Collapse of the rolling velocity data of fig. (3). Curve I plots $\ln(V)$ against $(1 - F/F_c)^2/K$ and curve II plots $\ln(VF_c/F)$ against $(1 - F/F_c)^{1.5}/K$. (b) Collapse of the same data when $\ln(V)$ is plotted against $(\sigma_1 + 1/K)[(1 - F/F_c)^{1.2}]$ with $\sigma_1$ = 108 s$^3$/m$^2$ and $F_c$ = 0.1mN. Similar symbols in figures (a) and (b) correspond to the same load.

With an overdamped system, the sphere could move by only one spacing length ($\lambda$) before it is pinned again. If we employ the latter scenario, the rolling velocity ($V=\lambda \nu$) would depend on $F$ and $K$ in the same way as does the rupture frequency. Thus, $V$ is given by:

$$V/V_o \approx \exp\left(-2U_b(1 - F/F_c)^n / mK\tau_L\right) \qquad (12)$$

Where $F$ is the applied bias. At this juncture, it would be prudent to point out that this form with n=1.5 is also consistent with the ball rolling (see Appendix) on a sinusoidal potential that is, perhaps, the simplest functional (or coarse grained) generalization of the rolling behavior accompanied by the pinning/depinning kinetics, excepting that the fibrillar detachments could lead to an avalanche (discussed in section 4) whereas rolling on a sinusoidal potential does not.

Various types of detachment modes are plausible as shown schematically in fig. (4). If the termini of the fibrils are truly flat ended, we expect that $\ln(V)$ to be proportional to $(1-F/F_c)^2/K$. When treated this way, the data do indeed cluster around a single curve, as was observed by us in a previous publication [45]. The bothersome feature here is that the overall rupture kinetics is non-Arrhenius, which is inconsistent with the observation that the rupture data exhibit an Arrhenius behavior over a significant range of the noise strength ($K$) at each applied bias. The direct observation of the fibril terminus using an optical profilometer (fig. (3b)) shows that it is, in fact, rounded with a radius of curvature ~ 40 $\mu$m. Thus it is more reasonable to try to collapse the data by plotting $\ln(V)$ against $(1-F/F_c)^{1.5}/K$. When attempted this way, good collapse of data (plot II of fig. (5a)) is obtained only when the drift velocity is divided by the bias. Although the curvature of the collapsed plot now is reduced from that of plot I, the overall rupture kinetics is still non-Arrhenius,.

There is however another angle from which to look at these data. Figure (3) reveals that all the $\ln V$ vs $1/K$ lines needed to fit the experimental data at all the biases, when extrapolated, tend to meet at a point farther to the left quadrant of the plot. A simple way to collapse the data would, therefore, be to first shift the $1/K$ axis to the right by a certain amount and then use this shifted values of $1/K$ to fit the data with an Arrhenius equation. Figure (5b) shows that this method works remarkably well. The idea of shifting the $1/K$ axis is equivalent to a generalized rupture kinetics of the form $V/V_o \sim \exp\left(-\dfrac{2(U_b+K\sigma)(1-F/F_c)^{1.2}}{mK\tau_L}\right)$. There are two issues related to this fit. The first of which is that the observed exponent (1.2) of $(1-F/F_c)$ is somewhat smaller than that (1.4 to 1.5) obtained from the simulations and secondly, the barrier energy needs to be modified by an additional entropy like term: $K\sigma$. In the context of a particle escaping from a potential well, Lin et al [41] suggested that an exponent of ~ 1 ensues when the applied force is much smaller than a critical force, which is clearly not the case in our current experiments. We believe that our results are influenced by other modes of separation of the fibrils, including peeling (fig. (4)) that being in a state of undifferentiated equilibrium [8] requires no activation. Postponing a detailed statistical analysis of this kind of mixed mode micro-rupture dynamics for future, we focus here on the other important issue related to the shift of the *1/K* axis that was required to collapse the experimental data.

## 4 Sequential rupture of fibrils

The basic premise here is that the fibrils do not detach all at once. When one fibril detaches from the surface, the load gets distributed to the remaining undetached fibrils thus enhancing the rupture rates of any of the remaining fibrils. The process continues till the load on the remaining fibrils are such that all of them detach ballistically, thus causing an avalanche. Within this scenario, the rupture kinetics may be described by the following equation:

$$\frac{d\phi}{dt} = -\phi v(\phi, P) \tag{13}$$

Where $\phi = \phi(t)$ is the fraction of the total numbers of fibrils that is in contact with the rigid sphere at any time *t*. For a spherical contact, $v(\phi, P)$ can be expressed as (see eq. (7)):

$$v(\phi, P) \approx v_o(\phi, P) \exp\left[-\frac{13.3(W^5 R^4 / E^{*2})^{1/3}(1 - P/\phi P_c)^{3/2}}{mK\tau_L / 2}\right] \tag{14}$$

The total time to rupture can be estimated by integrating equation 13 as follows:

$$T = \int_{P/P_c}^{1} \frac{d\phi}{\phi v(\phi, P)} \tag{15}$$

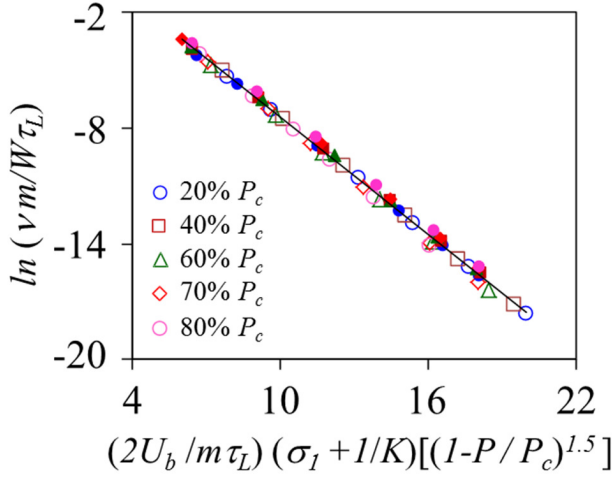

**Fig. 6.** (Colour on-line) Arrhenius plots of the frequency of detachment of multiple fibrils from a surface with a JKR contact. The parameters of these calculations are same as those of fig. (2), except that two different values of $W$ (0.04 J/m$^2$: open symbols; 0.01 J/m$^2$ filled symbols) were used. The data collapse in one master line when the normalized frequency $\ln(\nu m/W\tau_L)$ is plotted against $(2U_b/m\tau_L)(\sigma_1 + 1/K)[(1-P/P_c)^{1.5}]$ where $\sigma_1 = \sigma/U_b$ with the value of $\sigma$ as 48 pJ s$^3$/m$^2$.

By calculating the rupture frequencies ($1/T$) using eq. (13) to (15) for two different values of $W$, we attempted to collapse all the data as follows. First, the rupture frequency was normalized as $m\nu/W\tau_L$, where $W\tau_L/m$ is the characteristic escape frequency of mass $m$ fluctuating in the JKR potential (compare equations 2, 5 and 7). Next, we modulated $(1-P/P_c)^{1.5}/K$ with $2U_b/mK\tau_L$ so that the data obtained with different values of $W$ can be collapsed on to a single curve. With these normalizations, fig.(6) shows that $\ln(m\nu/W\tau_L)$ is indeed linear with $2U_b(1-P/P_c)^{1.5}/mK\tau_L$ provided that the horizontal axis is shifted by a constant amount. This analysis thus leads to an equation of the type shown below that provides partial justification for the shift of the $1/K$ axis of the experimental data as was done in fig. (5b).

$$\frac{m\nu}{W\tau_L} \approx \exp\left[-\frac{2(U_b + K\sigma)(1-P/P_c)^{1.5}}{mK\tau_L}\right] \qquad (15)$$

# 5 Discussions and final remarks

The main point of this paper is that the contact formed by the adhesive interaction of a soft deformable object with a rigid substrate can be broken sub-critically in the presence of a noise. This idea of the noise induced dissociation of a soft elastic contact has been useful in understanding some recently reported experimental results [45] of the pinning-de-pinning induced rolling of a rigid sphere on a soft fibrillar substrate. Although, there is a slight discrepancy in the exponent (1.2) of the reduced bias needed to fit the experimental data and that (1.4 – 1.5) expected of the detachment of a spherical contact, the discrepancy is not large. The kinetic analysis provided a new insight in that an "entropy" like term contributes to the energy barrier. Further studies are, however, required in analyzing the mixed mode ruptures of multi-fibrillar contacts in which load is shared by certain modes that are activated and others (i.e. peeling) that are not. Careful experiments with single fibrillar contacts with various other geometries are expected to provide further insights in these types of contact separation problems. The studies presented here could also be useful in understanding the pinning-depinning dynamics in various other types of bio-inspired adhesives and composites as well as understanding the (thermal) noise induced detachments of cells, macromolecules and soft colloids [52] from surfaces. Study of a noise induced separation of contact of soft materials may also be useful in obtaining the depth of the energy potential which may contrast and complement the conventional fracture mechanics methods of obtaining the strain energy release rates. We believe that noise induced detachments of soft adhesive contact may also find interesting applications in recently emerging transfer printing technologies [48].

## Appendix

**Langevin dynamics simulations of the splitting of soft contact**

The purpose of this section is to try to recover the result that the energy barrier to rupture a sphere from a rigid flat plate scales with the reduced bias as $(1-P/P_c)^{1.5}$ using a Langevin dynamics simulation. In order to accomplish this objective, our first step is to write down the

Lagrangian (*L*), in terms of the mass (*m*), elastic displacement (*δ*) and the energy of the system as

$$L = \frac{1}{2} m \dot{\delta}^2 - U(a) \tag{A1}$$

$$\delta = \frac{a^2}{3R}\left(1 + \frac{2PR}{E^* a^3}\right) \tag{A2}$$

*U*(*a*) is the thermodynamic potential energy, which is given by eq.(1). Now, solving the Lagrangian equation (eq. (A1)), we obtain the crack growth equation with a frictional dissipation as in eq. (A3).

$$\frac{\partial}{\partial t}\left(\frac{\partial L}{\partial \dot{a}}\right) = \frac{\partial L}{\partial a} - \frac{\partial \varepsilon}{\partial \dot{a}} \tag{A3}$$

Where $\varepsilon$ is the energy dissipation function $G\frac{dA}{dt}$, where $G\ (\sim \sigma_0^2 T\dot{a}/E^*)$ is the energy release rate in the linear friction regime [53] and $A=\pi a^2$. Here, *T* is the relaxation time of the adhering polymer chains, $\sigma_o$ is the cohesive stress, and *E*\* is the contact modulus. Now solving eq. (A3) and adding a noise term $\gamma(t)$, we have the eq. (8) of the text:

$$\left(\frac{2a}{3R} - \frac{2P}{3E^* a^2}\right)^2 \ddot{a} + \left(\frac{2a}{3R} - \frac{2P}{3E^* a^2}\right)\left(\frac{2}{3R} + \frac{4P}{3E^* a^3}\right)\dot{a}^2 + \frac{E^* a^4}{3mR^2} + \frac{P^2}{3mE^* a^2} - \frac{2Pa}{3mR} - \frac{2\pi Wa}{m} + \frac{4\pi a T \sigma_o^2}{mE^*}\dot{a} = \gamma(t) \tag{A4}$$

Although a more exact form of the friction is non-linear with the crack velocity [53], the linear friction model as used above is useful for capturing essential physics of the rupture dynamics that can be compared with a Kramers' model. In the current simulation, we treat the term $4\pi T\sigma_o^2/E^*$ of equation A4 as an empirical parameter in an overdamped limit. A computer generated [44] Gaussian random noise was used to integrate eq. (A4) using a fixed load condition. Logarithm of the rupture rate at each load was linear with $1/K$. All the rupture data can be collapsed (Fig. A1) by plotting $\ln(v/v_o)$ against the reduced bias as $(1-P/P_c)^{1.38}/K$. Note that that the exponent of the reduced bias is slightly smaller than 1.5.

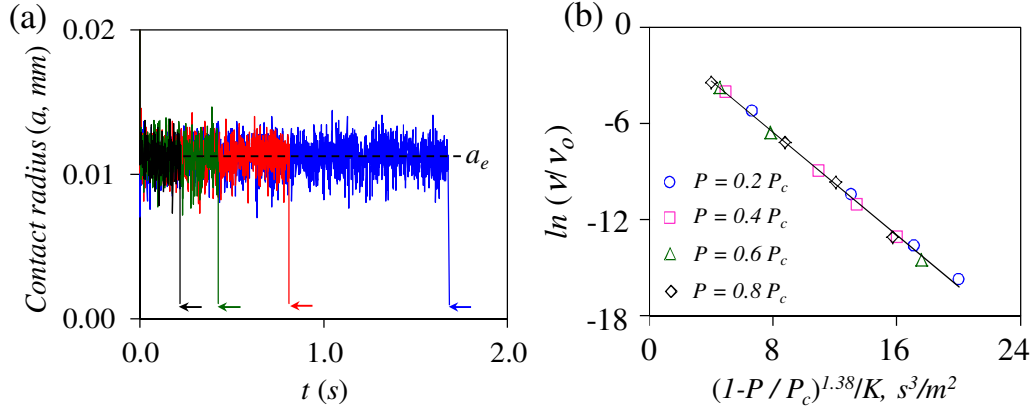

**Fig. A1.** (a) The fluctuation of the radius ($a$) of contact about a mean value ($a_e$) is obtained from the simulations based on eq. A4. The contact falls apart eventually (indicated by the arrows). From the mean value of the waiting times, a rupture frequency was estimated. (b) Summary of the rupture kinetics data using Langevin dynamics simulations (eq. (A4)). These calculations were made using the following parameters: $R= 100$ $\mu$m, $E^*= 1$MPa, $W=0.04$ J/m$^2$, $m=4.2$ $\mu$g with a friction term of eq. (A4) i.e. $(mE^*/4\pi\sigma_o^2 T)$ set as 12 ns.m.

**Motion over a periodic potential**

Motion of a particle over a periodic potential was used by Prandtl [22] to study the nature of friction. This model is also generic to study the motion of particle in a tilted potential [54]. Here, we consider a translational form of the stochastic rolling equation of motion of the sphere on a periodic potential of wavelength $\lambda$:

$$\frac{7}{5}\frac{dV}{dt} + \frac{V}{\tau_L} + \frac{\pi U_b}{m\lambda}\cos\frac{2\pi x}{\lambda} = \frac{F}{m} + \gamma(t), \tag{A5}$$

$$\frac{dx}{dt} = V, \tag{A6}$$

Here, $F$ ($=mg\sin\theta$) is the force acting through the center of gravity of the sphere parallel to the substrate. The force $P$ acting on each fibril is proportional to $F$ through a geometric factor.

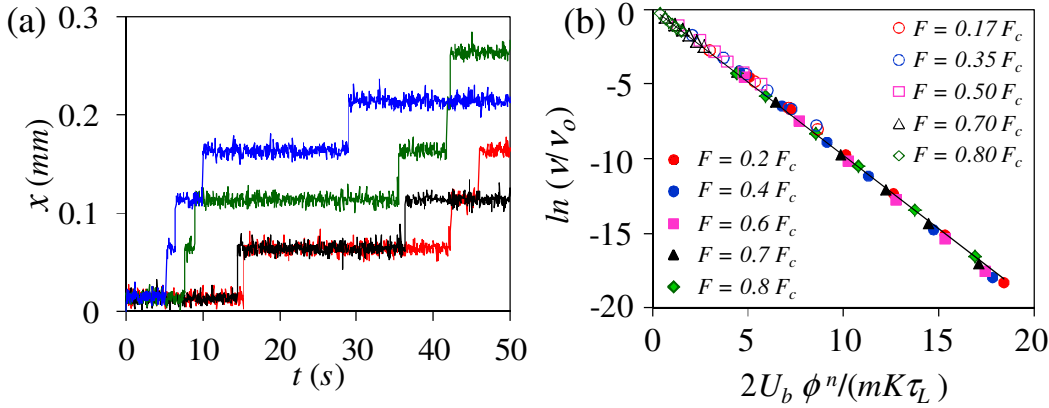

**Fig. A2**. (a) Typical trajectories of a sphere moving over a sinusoidal potential in the presence of a bias and an external noise. From the mean value of the waiting times, a barrier crossing frequency was estimated. (b) Comparison of rolling kinetics data as obtained from Langevin simulation (eq. (A5)) (open symbols) and Kramers' formalism (eq. (6)) (filled symbols). $\phi$ is the reduced bias. The value of $n$ is 1.4 for the Kramers' calculations and 1.5 for the rolling using Langevin dynamics. For the Kramers' calculations, the parameters are same as those of fig. (2), while for the Langevin dynamics simulations, the following parameters were used: $\lambda= 50$ $\mu$m, $\tau_L=0.001$ s, $h= 1.6$ $\mu$m, $U_b = 0.06$ pJ.

Using a computer generated Gaussian random noise, eq. (A5) and (A6) were integrated with a fixed value of the reduced bias: $\phi=1-F/F_c$. From the trajectories generated at each noise strength, the drift velocity was estimated directly. Logarithm of this drift velocity is linear with $1/K$ at each value of $\phi$. Now, the data collected at different values of $K$ and $\phi$ were normalized by plotting $\ln(V/V_o)$ against the reduced bias as $2U_b\phi^{1.5}/K$. This result was compared with the prediction of the fibril detachment model using Kramers theory in which $\ln(V/V_o)$ was plotted against $2U_b\phi^{1.4}/K$ with the value of $U_b$ as $13.3\dfrac{W^{5/3}R^{4/3}}{E^{*2/3}}$. Figure (A2) shows that the both set of data collapse on to a single curve thus demonstrating the rolling on a potential well is functionally equivalent to that accompanied with the detachment of fibrils.

Note Added in Proof

The current study considers the detachment kinetics in the strong friction limit. The effect of inertia in an underdamped limit deserves a detailed analysis, especially in understanding the avalanche dynamics associated with certain pinning-depinning transition, which will be accomplished in upcoming publication.